\documentclass[runningheads]{llncs}
\usepackage[acronym]{glossaries}
\usepackage{tikz}
\usetikzlibrary{positioning,arrows.meta,calc}
\usepackage{xcolor}
\usepackage{placeins}
\usepackage{esvect}
\makeglossaries
\usepackage[T1]{fontenc}
\usepackage{graphicx}
\begin{document}
%
\newacronym{uav}{UAV}{unmanned aerial vehicle}
\newacronym{milp}{MILP}{mixed-integer linear programming}
\newacronym{qubo}{QUBO}{quadratic unconstrained binary optimization}
\newacronym{sa}{SA}{simulated annealing}
\newacronym{qaoa}{QAOA}{Quantum Approximate Optimization Algorithm}
\title{Reducing QUBO Complexity in Drone Delivery Through Classical Feasibility Preprocessing}
%
%
\author{%
Silvie Ill\'{e}sov\'{a}\inst{4}\thanks{Corresponding author.} \and
Vojt\v{e}ch Nov\'{a}k\inst{1,2,3} \and
Martin Beseda\inst{5}}
\authorrunning{S. Ill\'{e}sov\'{a} et al.}
%
%
\institute{%
Department of Computer Science, Faculty of Electrical Engineering and Computer Science, 
VSB -- Technical University of Ostrava, Ostrava, Czech Republic
\and
IT4Innovations National Supercomputing Center, VSB -- Technical University of Ostrava,
708 00 Ostrava, Czech Republic
\and
Department of Informatics and Statistics, Marine Research Institute, Klaipeda University,
Klaipeda, Lithuania
\and
Gran Sasso Science Institute, L'Aquila, Italy\\
\email{silvie.illesova@gssi.it}
\and
Dipartimento di Ingegneria e Scienze dell'Informazione e Matematica,
Universit\`{a} dell'Aquila, Via Vetoio, I-67010 Coppito, L'Aquila, Italy}
\titlerunning{Reducing QUBO Complexity in Drone Delivery}
\maketitle              
\begin{abstract}
Drone-delivery planning combines routing decisions with operational constraints such as payload capacity and limited onboard energy, resulting in combinatorial optimization problems that can become costly to encode directly in binary form. This work investigates a hybrid quantum--classical approach in which candidate drone missions are generated and evaluated classically before constructing a quadratic unconstrained binary optimization (QUBO) model for mission selection. Payload- and energy-infeasible missions are removed during preprocessing, while the reduced QUBO is solved using exact enumeration, simulated annealing (SA), and the Quantum Approximate Optimization Algorithm (QAOA), with mixed-integer linear programming (MILP) used as a classical reference. Experiments on controlled synthetic delivery instances evaluate solution quality, the effect of preprocessing, and computational scaling. Exact QUBO optimization reproduces the MILP optimum on the evaluated small instances, while SA achieves $100\%$ feasibility with a median optimality gap of $0\%$. Combined payload-and-energy preprocessing reduces the mean number of QUBO variables by $29.60\%$ and quadratic interactions by $49.20\%$, while reducing mean SA runtime by $31.26\%$. In contrast, the evaluated shallow statevector-based QAOA configuration achieves lower feasibility and is restricted to small QUBO instances. These results indicate that, in the considered setting, the main practical benefit of the hybrid architecture comes from classical feasibility preprocessing that reduces the binary search space before QUBO-based optimization.

\end{abstract}
\section{Introduction}
\label{sec:introduction}

Currently, \glspl{uav} are increasingly considered for last-mile delivery because they can provide direct point-to-point transport, reduce dependence on road infrastructure, and support rapid delivery in areas where conventional ground transportation may be less efficient \cite{garg2023drones,shuaibu2025review,eskandaripour2023last}. At the same time, practical drone-delivery planning requires the simultaneous consideration of several operational constraints, including payload capacity, limited onboard energy, route length, fleet size, and the assignment of customer requests to feasible drone missions \cite{raivi2023drone,attenni2023drone,benarbia2021literature}. Together, these factors lead to combinatorial optimization problems whose complexity can increase rapidly with the number of customers and candidate routes.

Classical approaches to drone-delivery optimization commonly rely on \gls{milp} \cite{chen2025probability,thomas2024collaborative,song2018persistent}, vehicle-routing formulations \cite{euchi2021hybrid,dorling2016vehicle,thibbotuwawa2020unmanned}, decomposition methods \cite{mara2026decomposition,yurek2018decomposition}, and metaheuristic optimization \cite{mirzapour2024optimizing,gunay2023hybrid,almuhaideb2021optimization}. These approaches remain effective for many problem instances, but recent work has also explored quantum and quantum-inspired optimization techniques for routing, scheduling, and resource-allocation problems \cite{nguyen2026quacod,osaba2025solving,bassaleh2025quantum,holliday2025quadro}. In particular, binary optimization models provide a natural connection between logistics problems and emerging quantum optimization algorithms.

One widely used formulation in this context is \gls{qubo}. A \gls{qubo} problem can be expressed as
\begin{equation}
    \min_{\mathbf{x}\in\{0,1\}^{n}} \mathbf{x}^{\top} Q \mathbf{x},
    \label{eq:qubo-general}
\end{equation}
where the binary vector $\mathbf{x}$ represents optimization decisions and the matrix $Q$ encodes objective terms together with penalties associated with problem constraints. Such formulations can be addressed using classical heuristics, including \gls{sa} \cite{bertsimas1993simulated}, as well as quantum optimization methods such as \gls{qaoa} \cite{zhou2020quantum}.

A direct encoding of a complete drone-delivery problem as a \gls{qubo} \cite{glover2018tutorial}, however, can become large quickly. Binary variables may be required to represent route choices, customer assignments, vehicle usage, and auxiliary constraint terms. This is particularly relevant for current quantum computing approaches, where the size of the optimization model directly affects the number of required qubits or, in simulation, the computational cost of representing and evolving the corresponding quantum state.

For this reason, this work investigates a hybrid quantum--classical decomposition rather than encoding the complete drone-delivery problem directly into a single \gls{qubo}. In the proposed approach, classical preprocessing is used to generate candidate drone missions and evaluate their operational feasibility. Routes that violate payload or energy constraints are removed before the binary optimization model is constructed. The remaining mission-selection problem is then represented as a \gls{qubo}, where binary decision variables determine which feasible missions are selected to serve the customer requests.

The motivation for this architecture is to separate deterministic feasibility checks from the combinatorial selection problem. Payload and energy constraints can be evaluated efficiently in a classical preprocessing stage, while the reduced candidate set is subsequently passed to a \gls{qubo} solver. This decomposition has the potential to reduce both the dimensionality and the connectivity of the resulting optimization problem while preserving the relevant feasible solution space.

The experimental evaluation is divided into three research questions. \textbf{RQ1} asks how effectively \gls{qubo}-based optimization can solve energy-constrained drone-delivery planning problems compared with classical optimization. \textbf{RQ2} examines how classical feasibility preprocessing affects the size and solution quality of the resulting \gls{qubo} formulation. \textbf{RQ3} investigates how the computational behavior of the proposed approach changes as the delivery instance and resulting \gls{qubo} grow.

To answer these questions, the proposed framework is evaluated on controlled synthetic drone-delivery instances. A \gls{milp} formulation is used as the classical reference, while the corresponding \gls{qubo} instances are solved using exact enumeration for sufficiently small problems, \gls{sa}, and \gls{qaoa}. The evaluation considers solution feasibility, optimality gap, \gls{qubo} size, the number of quadratic interactions, and solver runtime.

The experiments are designed to examine both the correctness of the \gls{qubo} representation and the practical effect of classical preprocessing. In particular, the comparison between the \gls{milp} and exact \gls{qubo} solutions is used to verify whether the binary reformulation preserves the reference optimum on the considered small instances. The preprocessing experiments then quantify how removing payload- and energy-infeasible missions changes the size and connectivity of the resulting optimization problem, as well as the computational behavior of the selected solvers.

This work makes four main contributions. First, it introduces a hybrid quantum--classical formulation for energy-constrained multi-drone delivery in which candidate missions are generated and filtered classically before a \gls{qubo}-based mission-selection stage. Second, it evaluates the effect of payload and energy preprocessing on \gls{qubo} size, connectivity, runtime, and solution quality. Third, it provides a comparative evaluation of \gls{milp}, exact \gls{qubo} optimization, \gls{sa}, and \gls{qaoa} on a common set of drone-delivery instances. Finally, it provides a reproducible experimental framework for studying the behavior and current limitations of hybrid quantum--classical optimization for constrained drone-delivery planning.

The remainder of the paper is organized as follows. Section~\ref{sec:background} presents the background and related work. Section~\ref{sec:problem} introduces the drone-delivery model and optimization problem. Section~\ref{sec:method} describes the proposed hybrid quantum--classical approach and the corresponding \gls{qubo} formulation. Section~\ref{sec:experimental} presents the experimental design. Section~\ref{sec:results} reports the results for the three research questions. Section~\ref{sec:discussion} discusses the implications of the findings, while Section~\ref{sec:threats} addresses threats to validity. Finally, Section~\ref{sec:conclusion} concludes the paper.

\section{Background and Related Work}
\label{sec:background}

This section summarizes the background required for the proposed hybrid quantum--classical approach. It first discusses the optimization characteristics of drone-delivery systems, with particular emphasis on routing, payload, and energy constraints. It then introduces \gls{qubo} formulations and the main classical and quantum approaches used to solve them. Finally, related work on quantum and hybrid optimization for routing and logistics is reviewed to position the contribution of this study.

\subsection{Drone-Delivery Optimization}
\label{subsec:drone-delivery}

Drone-delivery planning can be formulated as a variant of vehicle-routing and resource-allocation problems in which conventional routing decisions are combined with physical and operational constraints specific to \glspl{uav} \cite{benarbia2021literature}. A typical planning problem must determine which customers are assigned to individual missions, the order in which they are visited, and how the resulting missions are distributed across the available fleet. Depending on the application, additional requirements may include delivery time windows, maximum mission duration, charging or battery replacement, heterogeneous vehicles, and restrictions on accessible flight space \cite{betti2024uav,seidakhmetov2022drone}.

Energy consumption is particularly important in drone-delivery optimization. Unlike many conventional routing models, where travel cost can be approximated mainly by distance, the energy consumption of a \gls{uav} may depend on both the traveled distance and the payload carried during each route segment \cite{zhang2021energy,beigi2022overview}. As deliveries are completed, the carried payload decreases, which means that the order in which customers are visited can affect the total energy required by a mission. Consequently, a route that is short in geometric terms is not necessarily the route with the lowest energy consumption.

These characteristics motivate energy-aware routing models in which battery capacity is treated explicitly as a feasibility constraint \cite{kim2026adaptive,choi2017optimization}. Given a route $r=(0,i_1,i_2,\ldots,i_k,0)$, where $0$ denotes the depot, a feasible mission must satisfy both the maximum payload capacity of the drone and the available onboard energy. A route may therefore be geometrically valid but operationally infeasible if the combined package weight exceeds the payload limit or if the estimated energy demand of the complete mission exceeds the available battery capacity.

The interaction between routing, payload, and energy constraints increases the size of the combinatorial search space. Even when each drone serves only a small number of customers per mission, the number of possible customer subsets and visit orders grows rapidly with the number of delivery requests. Exact mathematical programming approaches, including \gls{milp}, can provide optimal solutions for many problem instances and are commonly used as reference methods \cite{mahmoodi2026sustainable}. Larger or more complex variants are also addressed using decomposition techniques, local search, evolutionary algorithms, \gls{sa}, and other metaheuristic methods.

An important modelling decision is whether all operational constraints should be represented directly in the main optimization model or whether part of the problem can be handled before the optimization stage. The latter is particularly relevant when many theoretically possible routes are physically infeasible. In such cases, deterministic feasibility checks can eliminate unsuitable route candidates before the combinatorial optimization is performed, thereby reducing the size of the subsequent search space.

\subsection{QUBO Formulations}
\label{subsec:qubo}

\Gls{qubo} provides a general representation for binary combinatorial optimization problems. In its standard form, a \gls{qubo} problem is written as
\begin{equation}
    \min_{\mathbf{x}\in\{0,1\}^{n}} \mathbf{x}^{\top}Q\mathbf{x},
    \label{eq:background-qubo}
\end{equation}
where $\mathbf{x}$ is a vector of binary decision variables and $Q$ is a real-valued matrix containing the linear and quadratic coefficients of the objective function. Diagonal entries of $Q$ correspond to linear contributions, while off-diagonal entries encode interactions between pairs of binary variables.

Constrained optimization problems can be transformed into \gls{qubo} form by incorporating their constraints into the objective function through penalty terms \cite{ayodele2022penalty}. For an equality constraint of the form
\begin{equation}
    \sum_j a_j x_j=b,
\end{equation}
a corresponding quadratic penalty can be written as
\begin{equation}
    A\left(\sum_j a_jx_j-b\right)^2,
    \label{eq:qubo-penalty}
\end{equation}
where $A>0$ controls the strength of the constraint relative to the original objective. Because the decision variables are binary, the squared expression can be expanded into linear and quadratic terms and incorporated directly into the \gls{qubo} matrix.

The selection of penalty coefficients is an important modelling consideration \cite{de2026optimized,verma2022penalty}. If a penalty is too small, infeasible solutions may receive a lower objective value than valid solutions. If it is excessively large, the constraint terms can dominate the original objective and make heuristic optimization more difficult. The \gls{qubo} formulation must therefore preserve the intended constrained optimum while maintaining a numerical structure that remains suitable for the selected solver.

The difficulty of a \gls{qubo} problem depends not only on the number of binary variables but also on the number of quadratic interactions. A model containing $n$ binary variables can contain up to
\begin{equation}
    \frac{n(n-1)}{2}
\end{equation}
pairwise interactions in addition to its linear terms. Consequently, preprocessing that removes variables can also substantially reduce the connectivity and density of the resulting optimization model.

An important advantage of \gls{qubo} is that the formulation itself is independent of the optimization backend. Small instances can be solved by exact enumeration, while larger instances can be approached using classical heuristics such as \gls{sa} \cite{kagawa2020fully,lang2022strategic}. The same representation is also compatible with quantum optimization methods, allowing a common problem formulation to be used for comparative evaluation of classical and quantum solvers.

\subsection{Quantum Optimization and QAOA}
\label{subsec:qaoa}

Quantum optimization investigates whether quantum computational mechanisms can be used to solve difficult combinatorial optimization problems. Among methods designed for gate-based quantum computers, \gls{qaoa} is one of the most widely studied approaches for binary optimization \cite{azfar2025quantum,schmidbauer2024polynomial}. It is a variational quantum algorithm that alternates between evolution under a problem-dependent cost Hamiltonian and a mixing Hamiltonian. Related variational quantum algorithms similarly combine parameterized quantum circuits with classical optimization loops across a range of application domains \cite{beseda2024state,PhysRevA.111.022437,saoovqebenchmarking}.

For a \gls{qubo} problem, a binary decision variable can be mapped to a qubit operator using
\begin{equation}
x_i=\frac{1-Z_i}{2},
\label{eq:binary-pauli}
\end{equation}
where $Z_i$ is the Pauli-$Z$ operator acting on qubit $i$. Substituting this relation into the \gls{qubo} objective yields a cost Hamiltonian of the form
\begin{equation}
H_C=\sum_i h_iZ_i+\sum_{i<j}J_{ij}Z_iZ_j+c,
\label{eq:cost-hamiltonian}
\end{equation}
where $h_i$, $J_{ij}$, and $c$ are obtained from the coefficients of the original \gls{qubo} formulation.

At depth $p$, the \gls{qaoa} variational state can be written as
\begin{equation}
|\psi(\boldsymbol{\gamma},\boldsymbol{\beta})\rangle=
\prod_{\ell=1}^{p}
e^{-i\beta_{\ell}H_M}
e^{-i\gamma_{\ell}H_C}
|+\rangle^{\otimes n},
\label{eq:qaoa-state}
\end{equation}
where $H_M$ denotes the mixing Hamiltonian and $\boldsymbol{\gamma}$ and $\boldsymbol{\beta}$ are variational parameters. A classical optimizer updates these parameters based on objective estimates obtained from the quantum circuit, making \gls{qaoa} inherently hybrid in its execution. The choice of classical optimization strategy can substantially affect the convergence and robustness of variational quantum algorithms, particularly under sampling and hardware noise \cite{illesova2025numerical,saoovqebenchmarking,novak2025reliable,bezdvek2025classical,novak2026optimization}.

Increasing the circuit depth $p$ increases the expressiveness of the variational ansatz but also raises the number of parameters and the computational cost \cite{majumdar2021optimizing,schmidbauer2024polynomial}. On current quantum hardware, deeper circuits are additionally affected by accumulated noise, while optimizer choice, parameterization, and circuit configuration can further influence the observed performance of variational methods \cite{illesova2025qmetric,illesova2026importance}. In statevector simulation, the main limitation is instead the exponential scaling of the quantum state representation, which requires storing $2^n$ complex amplitudes for $n$ qubits. As a result, simulation-based \gls{qaoa} studies are generally restricted to relatively small \gls{qubo} instances.

\gls{qaoa} has also been investigated for discrete optimization problems outside routing and logistics, including software-engineering tasks such as test-case selection \cite{trovato2025preliminary}. This broader use of \gls{qaoa} illustrates its applicability to binary combinatorial problems beyond transportation-specific formulations.

This limitation is particularly relevant to routing and logistics problems, where direct encodings may require binary variables for route selection, customer assignment, vehicle usage, and additional constraints. Reducing the number of decision variables before the quantum optimization stage is therefore one possible strategy for extending the range of instances that can be investigated.

\subsection{Hybrid Quantum--Classical Optimization for Routing and Logistics}
\label{subsec:hybrid-related}

Routing, scheduling, packing, and resource-allocation problems have increasingly been investigated using quantum and quantum-inspired optimization techniques \cite{osaba2022systematic,gautam2023quantum}. Vehicle-routing variants are especially suitable for this research because many of their decisions can be represented using binary variables, enabling their objectives and constraints to be expressed as \gls{qubo} or Ising models.

Previous work has considered quantum annealing and gate-based quantum methods for vehicle routing, traveling-salesperson variants, scheduling, and logistics optimization \cite{gachnang2022quantum,phillipson2024quantum}. More recent studies have also examined drone-specific applications in which \gls{qubo}-based formulations represent routing, packing, or scheduling decisions under operational constraints \cite{kwa2025hybrid}. These studies demonstrate that quantum optimization models can be applied to drone logistics, but they also expose a recurring limitation: direct formulations can require a large number of binary variables even for relatively small delivery instances.

Hybrid approaches attempt to mitigate this limitation by assigning different parts of the workflow to classical and quantum components \cite{illesova2025classical}. Classical computation can be used for tasks that are deterministic, inexpensive, or inefficient to represent with binary variables, while the reduced combinatorial component is delegated to the quantum or quantum-inspired solver. Common examples include problem decomposition, candidate generation, clustering, constraint preprocessing, and classical post-processing of quantum solutions.

The approach investigated in this work follows this general principle but focuses specifically on feasibility preprocessing for drone missions. Candidate missions are generated classically, after which payload and energy constraints can be evaluated before constructing the \gls{qubo}. The optimization layer therefore operates on mission-selection decisions rather than attempting to represent every physical constraint directly inside the binary model.

This separation makes it possible to evaluate the role of the classical component explicitly. By constructing \gls{qubo} formulations with different levels of preprocessing, the experiments can quantify how classical filtering affects the number of binary variables, the number of quadratic interactions, solver runtime, feasibility, and solution quality. The resulting evaluation therefore addresses not only the performance of individual solvers but also the practical effect of decomposing drone-delivery optimization into classical feasibility analysis and \gls{qubo}-based combinatorial selection.

\section{Problem Formulation}
\label{sec:problem}

The considered optimization problem concerns the selection of feasible drone-delivery missions for serving a set of customer requests from a common depot. The formulation separates the physical feasibility of individual missions from the combinatorial decision of which missions should be selected. This separation is later used by the proposed hybrid quantum--classical method, where mission generation and feasibility checking are performed classically and the resulting mission-selection problem is represented as a \gls{qubo}.

\subsection{Delivery Scenario and Notation}
\label{subsec:scenario}

Let the set of customers be denoted by $\mathcal{V}=\{1,\ldots,n\}$ and let node $0$ represent the depot. Each customer $i\in\mathcal{V}$ is associated with a spatial position $(x_i,y_i)$ and a package weight $w_i$. The depot is similarly associated with a fixed position $(x_0,y_0)$. The Euclidean distance between any two locations $i$ and $j$ is denoted by $d_{ij}$ and is computed as
\begin{equation}
    d_{ij}=\sqrt{(x_i-x_j)^2+(y_i-y_j)^2}.
    \label{eq:distance}
\end{equation}

A drone mission is defined as an ordered route
\begin{equation}
    r=(0,i_1,i_2,\ldots,i_k,0),
    \label{eq:mission}
\end{equation}
where $i_1,\ldots,i_k$ are distinct customers served by the mission. Each mission therefore starts and ends at the depot and may contain one or more delivery stops. The complete set of generated candidate missions is denoted by $\mathcal{R}$.

For each mission $r\in\mathcal{R}$, the total traveled distance is
\begin{equation}
    D_r=\sum_{(i,j)\in r} d_{ij},
    \label{eq:route-distance}
\end{equation}
where the summation covers all consecutive edges of the route, including the departure from and return to the depot.

The binary parameter $a_{ir}$ indicates whether customer $i$ is served by mission $r$,
\begin{equation}
    a_{ir}=
    \begin{cases}
        1, & \text{if customer } i \text{ is included in mission } r,\\
        0, & \text{otherwise.}
    \end{cases}
    \label{eq:route-membership}
\end{equation}

\subsection{Drone Mission Feasibility}
\label{subsec:feasibility}

Each drone is characterized by a maximum payload capacity $W_{\max}$ and an available energy capacity $E_{\max}$. A candidate mission is considered operationally feasible only if it satisfies both constraints.

The total initial payload of mission $r$ is given by
\begin{equation}
    W_r=\sum_{i\in r} w_i.
    \label{eq:route-payload}
\end{equation}
The payload-feasibility condition is therefore
\begin{equation}
    W_r\leq W_{\max}.
    \label{eq:payload-constraint}
\end{equation}

Energy consumption is modeled as dependent on both traveled distance and the payload carried during each segment. Let $W_{ij}^{(r)}$ denote the payload carried by the drone while traversing edge $(i,j)$ in mission $r$. The energy consumed on that edge is modeled as
\begin{equation}
    E_{ij}^{(r)} = d_{ij}\left(e_0+e_1W_{ij}^{(r)}\right),
    \label{eq:edge-energy}
\end{equation}
where $e_0$ represents the distance-dependent energy consumption of the unloaded vehicle and $e_1$ controls the additional energy demand associated with payload.

Because packages are removed from the vehicle as deliveries are completed, the payload generally decreases throughout a mission. The total energy required by route $r$ is therefore
\begin{equation}
    E_r=\sum_{(i,j)\in r}E_{ij}^{(r)}.
    \label{eq:route-energy}
\end{equation}
The energy-feasibility condition is
\begin{equation}
    E_r\leq E_{\max}.
    \label{eq:energy-constraint}
\end{equation}

A mission is retained in the feasible candidate set only if all relevant operational conditions are satisfied. Depending on the preprocessing configuration used in the experiments, either no feasibility filtering, payload-only filtering, or combined payload-and-energy filtering is applied before the optimization model is constructed.

\subsection{Optimization Objective}
\label{subsec:objective}

After candidate missions have been generated, the remaining optimization problem is to select a subset of missions that serves every customer while minimizing the total delivery cost. For each candidate mission $r\in\mathcal{R}$, let the binary decision variable
\begin{equation}
    x_r=
    \begin{cases}
        1, & \text{if mission } r \text{ is selected},\\
        0, & \text{otherwise}
    \end{cases}
    \label{eq:route-variable}
\end{equation}
indicate whether the mission is included in the final delivery plan.

The cost associated with mission $r$ is defined as
\begin{equation}
    C_r=\alpha D_r+\beta E_r+\gamma,
    \label{eq:route-cost}
\end{equation}
where $\alpha$ and $\beta$ weight the contributions of travel distance and energy consumption, respectively, while $\gamma$ represents a fixed cost associated with activating an additional mission. The fixed term can be interpreted as a simple proxy for fleet utilization or mission overhead.

The corresponding constrained mission-selection problem is
\begin{equation}
    \min_{\mathbf{x}} \sum_{r\in\mathcal{R}} C_r x_r
    \label{eq:mission-objective}
\end{equation}
subject to
\begin{equation}
    \sum_{r\in\mathcal{R}} a_{ir}x_r = 1,
    \qquad \forall i\in\mathcal{V},
    \label{eq:customer-coverage}
\end{equation}
and
\begin{equation}
    x_r\in\{0,1\},
    \qquad \forall r\in\mathcal{R}.
    \label{eq:binary-route}
\end{equation}

Constraint~\eqref{eq:customer-coverage} requires every customer to be served exactly once. This prevents both unserved requests and duplicate delivery assignments. Because route feasibility is handled during the candidate-generation stage, the mission-selection model itself does not need to encode payload and energy constraints directly when feasibility preprocessing is applied.

The same mission-selection problem can be represented as a \gls{qubo} by incorporating the coverage constraints into the objective through a quadratic penalty. Using penalty coefficient $A>0$, the unconstrained objective becomes
\begin{equation}
    H(\mathbf{x})=
    \sum_{r\in\mathcal{R}} C_r x_r
    +
    A\sum_{i\in\mathcal{V}}
    \left(
        1-\sum_{r\in\mathcal{R}} a_{ir}x_r
    \right)^2.
    \label{eq:mission-qubo}
\end{equation}

The first term of Eq.~\eqref{eq:mission-qubo} minimizes the total cost of the selected missions, while the second penalizes any solution in which a customer is served either zero times or more than once. Since $x_r^2=x_r$ for binary variables, the penalty expression can be expanded into linear and quadratic terms, yielding a standard \gls{qubo} representation.

The formulation in Eq.~\eqref{eq:mission-qubo} is the optimization model used by the exact \gls{qubo}, \gls{sa}, and \gls{qaoa} solvers in the experimental evaluation. The corresponding constrained formulation in Eqs.~\eqref{eq:mission-objective}--\eqref{eq:binary-route} is solved using \gls{milp} and serves as the classical reference solution.

\section{Hybrid Quantum--Classical Method}
\label{sec:method}

The proposed method separates the drone-delivery optimization workflow into a classical feasibility stage and a binary mission-selection stage. The classical component generates candidate missions, evaluates their payload and energy requirements, and removes missions that violate the selected operational constraints. The remaining mission-selection problem is then transformed into a \gls{qubo} and solved using either classical or quantum optimization methods. This decomposition is intended to reduce the size and connectivity of the binary optimization problem before it is passed to the \gls{qubo} solver.

\subsection{Overview of the Hybrid Workflow}
\label{subsec:workflow}

The workflow begins from a drone-delivery instance containing the depot position, customer locations, package weights, and drone operating parameters. Candidate missions are first generated classically from subsets and permutations of the customer set. Each candidate mission is evaluated in terms of route distance, carried payload, and energy consumption. Depending on the preprocessing configuration, missions that violate payload or energy limits are discarded. The remaining missions define the binary decision space of the optimization problem.

For each retained mission, a binary variable indicates whether that mission is selected. The mission costs and customer-coverage relations are then used to construct the \gls{qubo} defined in Eq.~\eqref{eq:mission-qubo}. The resulting binary optimization problem can be solved using exact enumeration, \gls{sa}, or \gls{qaoa}. The selected binary solution is finally decoded into a set of drone missions, after which customer coverage and operational feasibility are verified classically.

The complete workflow can therefore be summarized as
\begin{equation}
    \text{instance generation}
    \rightarrow
    \text{candidate missions}
    \rightarrow
    \text{feasibility preprocessing}
    \rightarrow
    \text{\gls{qubo}}
    \rightarrow
    \text{solver}
    \rightarrow
    \text{decoded delivery plan}.
    \label{eq:workflow}
\end{equation}

\subsection{Classical Candidate-Mission Generation}
\label{subsec:mission-generation}

Candidate missions are generated before the binary optimization stage. For a delivery instance with customer set $\mathcal{V}$, the generator constructs missions containing one or more customers, subject to a configurable maximum number of stops per mission. For a subset $\{i_1,\ldots,i_k\}\subseteq\mathcal{V}$, different customer orders correspond to different missions because both route distance and payload-dependent energy consumption can depend on the visit sequence.

Each generated mission has the form
\begin{equation}
    r=(0,i_1,\ldots,i_k,0),
\end{equation}
and is evaluated by computing the total route distance $D_r$, initial payload $W_r$, and total energy requirement $E_r$. The customer-membership coefficients $a_{ir}$ are also recorded during this stage and later used to construct the coverage penalties in the \gls{qubo}.

For small instances, the candidate set can be generated exhaustively up to the configured maximum route length. As the number of customers increases, exhaustive route construction becomes impractical because the number of subsets and permutations grows rapidly. The implementation therefore allows the candidate set to be capped for larger instances. This cap limits the dimension of the subsequent \gls{qubo} and enables controlled scaling experiments, although it also means that customer count alone is not a direct measure of the resulting binary problem size.

\subsection{Feasibility Preprocessing}
\label{subsec:preprocessing}

The classical preprocessing stage evaluates the physical feasibility of every generated mission before the \gls{qubo} is constructed. Three preprocessing configurations are considered in the experimental evaluation: no filtering, payload filtering, and combined payload-and-energy filtering.

In the unfiltered configuration, all generated candidate missions are retained regardless of their physical feasibility. This configuration provides a reference for quantifying the reduction obtained through classical preprocessing.

With payload filtering, a mission $r$ is retained only if
\begin{equation}
    W_r \leq W_{\max}.
    \label{eq:method-payload-filter}
\end{equation}
This eliminates missions whose combined package weight exceeds the carrying capacity of the drone.

With combined payload-and-energy filtering, the payload condition in Eq.~\eqref{eq:method-payload-filter} is applied together with
\begin{equation}
    E_r \leq E_{\max}.
    \label{eq:method-energy-filter}
\end{equation}
The retained mission set is therefore
\begin{equation}
    \mathcal{R}_{\mathrm{feasible}}
    =
    \left\{
        r\in\mathcal{R}
        \mid
        W_r\leq W_{\max}
        \land
        E_r\leq E_{\max}
    \right\}.
    \label{eq:feasible-route-set}
\end{equation}

The purpose of this stage is not to solve the global routing problem classically, but to remove individual missions that cannot appear in any operationally valid solution. Consequently, the preprocessing stage reduces the number of binary decisions without directly deciding which combination of feasible missions should be selected.

\subsection{QUBO Mission-Selection Formulation}
\label{subsec:mission-selection}

After preprocessing, one binary variable $x_r$ is introduced for every retained candidate mission $r$. The linear contribution of each variable is determined by its mission cost $C_r$, while the customer-coverage constraints are converted into quadratic penalties.

Starting from
\begin{equation}
    H(\mathbf{x})
    =
    \sum_{r\in\mathcal{R}}C_rx_r
    +
    A\sum_{i\in\mathcal{V}}
    \left(
        1-\sum_{r\in\mathcal{R}}a_{ir}x_r
    \right)^2,
\end{equation}
the coverage term for a single customer $i$ can be expanded as
\begin{equation}
    \left(
        1-\sum_r a_{ir}x_r
    \right)^2
    =
    1
    -
    2\sum_r a_{ir}x_r
    +
    \left(
        \sum_r a_{ir}x_r
    \right)^2.
\end{equation}

Using $x_r^2=x_r$, the final term becomes
\begin{equation}
    \left(
        \sum_r a_{ir}x_r
    \right)^2
    =
    \sum_r a_{ir}x_r
    +
    2\sum_{r<s}a_{ir}a_{is}x_rx_s.
\end{equation}

The resulting \gls{qubo} therefore contains linear coefficients associated with mission cost and customer coverage, together with quadratic coefficients coupling pairs of missions that serve the same customer. Selecting two overlapping missions increases the objective through these quadratic penalties, while leaving a customer uncovered also incurs a penalty.

The number of \gls{qubo} variables is equal to the number of retained candidate missions,
\begin{equation}
    n_{\mathrm{QUBO}} = |\mathcal{R}|,
    \label{eq:qubo-variable-count}
\end{equation}
while the number of quadratic interactions depends on how many candidate missions overlap in the customers they contain. Classical preprocessing can therefore reduce both $n_{\mathrm{QUBO}}$ and the interaction count by removing infeasible missions before the matrix $Q$ is assembled.

\subsection{Classical and Quantum Solution Paths}
\label{subsec:solution-paths}

The same mission-selection formulation is evaluated through multiple solver paths. The constrained formulation is solved using \gls{milp} and provides the reference objective value. For sufficiently small \gls{qubo} instances, exact enumeration is used to identify the global minimum of the binary model and to verify that the reformulation preserves the \gls{milp} optimum.

\Gls{sa} is used as a classical heuristic \gls{qubo} solver. Starting from an initial binary solution, the algorithm repeatedly proposes local state changes and accepts them according to an annealing schedule that gradually reduces the probability of accepting higher-energy states \cite{tambunan2026quantum,suzuki2025simulated}. This provides a scalable classical baseline operating directly on the same \gls{qubo} representation as the quantum solver.

For the quantum path, the \gls{qubo} coefficients are mapped to the cost Hamiltonian described in Eq.~\eqref{eq:cost-hamiltonian} and optimized using \gls{qaoa}. The implementation uses a statevector simulator, which allows the behavior of the variational algorithm to be studied without introducing hardware noise. Because statevector memory scales exponentially with the number of qubits, \gls{qaoa} is executed only when the resulting \gls{qubo} remains below the configured simulation limit.

\subsection{Solution Decoding}
\label{subsec:decoding}

Each solver ultimately produces a binary vector
\begin{equation}
    \mathbf{x}=(x_1,\ldots,x_{|\mathcal{R}|}),
\end{equation}
which is decoded by selecting every mission $r$ for which $x_r=1$. The decoded solution is then checked against the original delivery requirements.

Customer coverage is evaluated using
\begin{equation}
    c_i=\sum_{r\in\mathcal{R}}a_{ir}x_r.
    \label{eq:decoded-coverage}
\end{equation}
A decoded solution is considered coverage-feasible only if
\begin{equation}
    c_i=1,
    \qquad
    \forall i\in\mathcal{V}.
\end{equation}

When feasibility preprocessing is active, the selected missions have already satisfied the corresponding payload and energy checks individually. For the unfiltered configuration, selected missions are additionally evaluated against the original operational constraints during decoding. This distinction allows the experiments to measure whether classical preprocessing changes the frequency with which the binary solver returns a physically valid delivery plan.

For every decoded solution, the implementation records the objective value, total route distance, total energy consumption, number of selected missions, customer-coverage violations, operational feasibility, solver runtime, and the dimensions of the corresponding \gls{qubo}. These quantities form the basis of the evaluation described in Section~\ref{sec:experimental}.

\section{Experimental Design}
\label{sec:experimental}

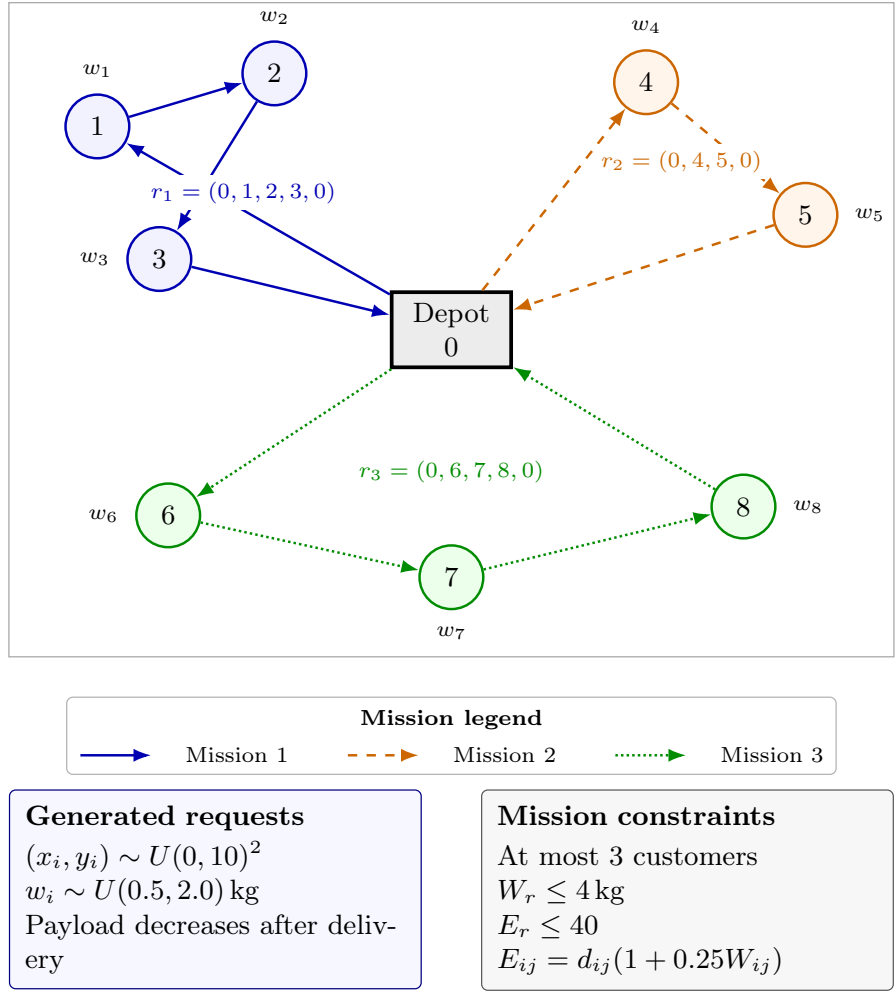
\begin{figure}[t]
\centering
\resizebox{0.96\linewidth}{!}{%
\begin{tikzpicture}[
    every node/.style={font=\small},
    depot/.style={
        rectangle,
        draw=black,
        fill=gray!15,
        very thick,
        minimum width=1.35cm,
        minimum height=0.85cm,
        align=center
    },
    customerA/.style={
        circle,
        draw=blue!70!black,
        fill=blue!5,
        thick,
        minimum size=0.72cm,
        inner sep=0pt
    },
    customerB/.style={
        circle,
        draw=orange!80!black,
        fill=orange!8,
        thick,
        minimum size=0.72cm,
        inner sep=0pt
    },
    customerC/.style={
        circle,
        draw=green!55!black,
        fill=green!8,
        thick,
        minimum size=0.72cm,
        inner sep=0pt
    },
    routeA/.style={
        -{Latex[length=2mm]},
        thick,
        draw=blue!70!black
    },
    routeB/.style={
        -{Latex[length=2mm]},
        thick,
        dashed,
        draw=orange!80!black
    },
    routeC/.style={
        -{Latex[length=2mm]},
        thick,
        densely dotted,
        draw=green!55!black
    },
    infoA/.style={
        rectangle,
        draw=blue!50!black,
        fill=blue!3,
        rounded corners=2pt,
        align=left,
        inner sep=5pt,
        font=\small
    },
    infoB/.style={
        rectangle,
        draw=gray!70!black,
        fill=gray!6,
        rounded corners=2pt,
        align=left,
        inner sep=5pt,
        font=\small
    },
    legendbox/.style={
        rectangle,
        draw=gray!60,
        fill=white,
        rounded corners=2pt,
        align=center,
        inner sep=4pt,
        font=\scriptsize
    }
]

\draw[thin, gray!70] (-5,-3.7) rectangle (5,3.7);

\node[
    anchor=south west,
    font=\small
] at (-5,3.80)
{$10\,\mathrm{km}\times10\,\mathrm{km}$ service region};

\node[depot] (depot) at (0,0) {Depot\\$0$};

\node[customerA] (c1) at (-4.0,2.3) {$1$};
\node[customerA] (c2) at (-2.0,2.9) {$2$};
\node[customerA] (c3) at (-3.3,0.8) {$3$};

\node[customerB] (c4) at (2.2,2.8) {$4$};
\node[customerB] (c5) at (4.0,1.3) {$5$};

\node[customerC] (c6) at (-3.2,-2.1) {$6$};
\node[customerC] (c7) at (0,-2.8) {$7$};
\node[customerC] (c8) at (3.3,-2.0) {$8$};

\node[above=2pt of c1, font=\scriptsize] {$w_1$};
\node[above=2pt of c2, font=\scriptsize] {$w_2$};
\node[left=2pt of c3, font=\scriptsize] {$w_3$};

\node[above=2pt of c4, font=\scriptsize] {$w_4$};
\node[right=2pt of c5, font=\scriptsize] {$w_5$};

\node[left=2pt of c6, font=\scriptsize] {$w_6$};
\node[below=2pt of c7, font=\scriptsize] {$w_7$};
\node[right=2pt of c8, font=\scriptsize] {$w_8$};

\draw[routeA] (depot) -- (c1);
\draw[routeA] (c1) -- (c2);
\draw[routeA] (c2) -- (c3);
\draw[routeA] (c3) -- (depot);

\draw[routeB] (depot) -- (c4);
\draw[routeB] (c4) -- (c5);
\draw[routeB] (c5) -- (depot);

\draw[routeC] (depot) -- (c6);
\draw[routeC] (c6) -- (c7);
\draw[routeC] (c7) -- (c8);
\draw[routeC] (c8) -- (depot);

\node[
    font=\scriptsize,
    fill=white,
    text=blue!70!black,
    inner sep=2pt
] at (-2.35,1.55)
{$r_1=(0,1,2,3,0)$};

\node[
    font=\scriptsize,
    fill=white,
    text=orange!80!black,
    inner sep=2pt
] at (2.6,1.9)
{$r_2=(0,4,5,0)$};

\node[
    font=\scriptsize,
    fill=white,
    text=green!55!black,
    inner sep=2pt
] at (0,-1.6)
{$r_3=(0,6,7,8,0)$};

\node[
    legendbox,
    anchor=north
] at (0,-4.15)
{
\textbf{Mission legend}\\[4pt]
\tikz{\draw[routeA] (0,0) -- (0.8,0);}
\quad Mission 1
\qquad
\tikz{\draw[routeB] (0,0) -- (0.8,0);}
\quad Mission 2
\qquad
\tikz{\draw[routeC] (0,0) -- (0.8,0);}
\quad Mission 3
};

\node[
    infoA,
    anchor=north west,
    text width=4.3cm
] at (-5,-5.2)
{
\textbf{Generated requests}\\[2pt]
$(x_i,y_i)\sim U(0,10)^2$\\
$w_i\sim U(0.5,2.0)\,\mathrm{kg}$\\
Payload decreases after delivery
};

\node[
    infoB,
    anchor=north east,
    text width=4.3cm
] at (5,-5.2)
{
\textbf{Mission constraints}\\[2pt]
At most $3$ customers\\
$W_r\leq4\,\mathrm{kg}$\\
$E_r\leq40$\\
$E_{ij}=d_{ij}(1+0.25W_{ij})$
};

\end{tikzpicture}%
}
\caption{Illustrative drone-delivery instance corresponding to the implemented model. Customers are generated within a $10\times10$ km service region around a central depot and are associated with package weights. Candidate missions start and end at the depot and contain at most three customer stops. Colors and line styles distinguish the example missions, while payload and energy feasibility are evaluated before construction of the \gls{qubo}.}
\label{fig:delivery-scenario}
\end{figure}

The experimental evaluation is designed to assess the proposed hybrid quantum--classical drone-delivery optimization approach from three complementary perspectives: solution quality, the effect of classical feasibility preprocessing, and computational scaling. Controlled synthetic delivery instances are used so that the number of customers and operational parameters can be varied systematically while keeping the evaluation reproducible. An illustrative delivery scenario, including the depot, generated customer requests, and example candidate missions, is shown in Fig.~\ref{fig:delivery-scenario}. Across the experiments, \gls{milp} is used as the classical reference, while the corresponding \gls{qubo} formulations are solved using exact enumeration for sufficiently small instances, \gls{sa}, and the \gls{qaoa}.

The evaluation considers both optimization quality and computational characteristics. Solution quality is assessed through feasibility and optimality gap with respect to the \gls{milp} reference solution, while the effect of preprocessing is measured through changes in the number of retained candidate missions, \gls{qubo} variables, and quadratic interactions. Solver runtime is additionally recorded to characterize computational behavior as the delivery instances and resulting \gls{qubo} formulations grow.

\FloatBarrier
\subsection{Research Questions}
\label{subsec:research-questions}

The experimental evaluation is organized around three research questions that examine complementary aspects of the proposed hybrid quantum--classical approach. The first focuses on solution quality, the second on the effect of classical preprocessing, and the third on computational scaling.

\paragraph{RQ1: Solution Quality.}
How effectively can \gls{qubo}-based optimization solve energy-constrained drone-delivery planning problems compared with classical optimization? The \gls{milp} formulation is used as the classical reference, while the corresponding mission-selection problem is solved using exact \gls{qubo} enumeration for sufficiently small instances, \gls{sa}, and \gls{qaoa}. The comparison considers feasibility and optimality gap with respect to the \gls{milp} reference solution. Agreement between the \gls{milp} and exact \gls{qubo} solutions is additionally used to verify that the binary reformulation preserves the reference optimum.

\paragraph{RQ2: Effect of Classical Preprocessing.}
How does classical feasibility preprocessing affect the size and solution quality of the resulting \gls{qubo} formulation? Three preprocessing configurations are compared: no filtering, payload-only filtering, and combined payload-and-energy filtering. Their effect is evaluated through the number of retained candidate missions, the number of \gls{qubo} variables, the number of quadratic interactions, solver runtime, feasibility, and optimality gap. This question examines whether inexpensive classical feasibility checks can reduce the combinatorial search space without removing the mission combination corresponding to the reference \gls{milp} optimum.

\paragraph{RQ3: Computational Scaling.}
How does the computational behavior of the proposed approach change as the delivery instance and resulting \gls{qubo} grow? The analysis considers both the number of customers and the resulting \gls{qubo} dimension because candidate-route generation and route capping mean that customer count is not necessarily proportional to the number of binary variables. Solver runtime is therefore evaluated against both customer count and \gls{qubo} size. For \gls{qaoa}, the analysis is restricted to instances that remain within the configured statevector simulation limit, while the classical methods are evaluated on larger instances.

\subsection{Synthetic Drone-Delivery Instances}
\label{subsec:instances}

The experiments use synthetically generated drone-delivery instances in order to control the number of customers and operational parameters while allowing multiple independent problem realizations to be evaluated. Each instance represents a square $10\times10$ km service region with a single depot located at its center. Customer locations are sampled independently from a uniform distribution over the service region, such that $(x_i,y_i)\sim U(0,10)^2$, and each customer is associated with a package weight sampled as $w_i\sim U(0.5,2.0)$ kg. An illustrative instance and the corresponding candidate missions are shown in Fig.~\ref{fig:delivery-scenario}.

Candidate missions start and end at the depot and contain at most three customers. Different visit orders are treated as distinct candidate missions because both traveled distance and payload-dependent energy consumption depend on the order in which deliveries are performed. For each mission, the initial payload is given by the sum of the package weights assigned to that mission, and the carried payload decreases after each delivery. When candidate-route capping is applied, all single-customer missions are retained to preserve the possibility of serving every customer, while the remaining available positions are filled by the lowest-cost multi-customer missions.

The maximum drone payload is set to $W_{\max}=4$ kg. Energy consumption on an edge $(i,j)$ is calculated according to the model introduced in Eq.~\eqref{eq:edge-energy}, using the experimental parameterization
\begin{equation}
    E_{ij}=d_{ij}\left(1+0.25W_{ij}\right),
    \label{eq:experimental-energy}
\end{equation}
where $W_{ij}$ denotes the payload carried while traversing the corresponding edge. The available mission energy is limited to $E_{\max}=40$. Consequently, candidate missions may be excluded during preprocessing because of either excessive payload or excessive estimated energy consumption.

The cost of mission $r$ is computed using the distance, energy, and fixed mission coefficients defined in Eq.~\eqref{eq:route-cost}. The \gls{qubo} coverage penalty is selected adaptively from the generated mission set as
\begin{equation}
    A = \lambda \max_{r\in\mathcal{R}} C_r,
    \label{eq:coverage-penalty-selection}
\end{equation}
where $\lambda$ denotes the configured coverage-penalty factor.

Synthetic generation is used instead of a fixed benchmark instance because the evaluation focuses on the behavior of the optimization formulation under controlled changes in problem size rather than on reproducing a particular real-world delivery network. Random seeds are fixed and recorded for each generated instance so that all solver configurations are evaluated on identical delivery scenarios. The main experimental settings are summarized in Table~\ref{tab:experimental-configuration}.

\begin{table}[t]
\centering
\caption{Main experimental configuration.}
\label{tab:experimental-configuration}
\begin{tabular}{ll}
\hline
\textbf{Parameter} & \textbf{Value} \\
\hline
Service region & $10\times10$ km \\
Depot location & Center of service region \\
Customer coordinates & $(x_i,y_i)\sim U(0,10)^2$ \\
Package weights & $w_i\sim U(0.5,2.0)$ kg \\
Maximum payload $W_{\max}$ & $4$ kg \\
Maximum mission energy $E_{\max}$ & $40$ \\
Maximum stops per mission & $3$ \\
Base energy coefficient $e_0$ & $1.0$ \\
Payload energy coefficient $e_1$ & $0.25$ \\
Distance weight $\alpha$ & $1.0$ \\
Energy weight $\beta$ & $0.20$ \\
Mission fixed cost $\gamma$ & $2.0$ \\
Coverage penalty factor $\lambda$ & $10.0$ \\
RQ1 customer counts & $4,5,6$ \\
RQ2 customer counts & $4,5,6,8$ \\
RQ3 customer counts & $4,6,8,10,12$ \\
Instances per problem size & $20$ \\
RQ1 stochastic repetitions & $10$ per instance \\
SA steps & $30{,}000$ \\
SA restarts & $3$ \\
Exact-QUBO limit & $24$ variables \\
QAOA maximum QUBO size & $16$ variables \\
QAOA depth $p$ & $1$ \\
QAOA optimizer & COBYLA \\
QAOA optimizer iterations & $80$ \\
QAOA samples & $1024$ \\
MILP time limit & $120$ s \\
\hline
\end{tabular}
\end{table}

\subsection{Compared Solvers}
\label{subsec:solvers}

Four optimization approaches are compared. The \gls{milp} formulation provides the classical reference solution and is solved using the SciPy mixed-integer linear programming interface with a time limit of $120$ s. Exact \gls{qubo} enumeration is additionally applied to instances containing at most $24$ binary variables in order to verify that the binary formulation reproduces the \gls{milp} optimum.

\Gls{sa} is used as a classical heuristic operating directly on the \gls{qubo} representation. Each run performs $30{,}000$ update steps and uses three independent restarts. The temperature follows an exponential schedule from an instance-dependent initial value to $10^{-3}$. For RQ1, \gls{sa} is repeated $10$ times per problem instance using different solver seeds.

The quantum optimization path uses \gls{qaoa} with a custom statevector simulator. The experiments use depth $p=1$ and restrict execution to \gls{qubo} instances containing at most $16$ binary variables. The variational parameters are initialized randomly, with $\gamma\sim U(0,\pi)$ and $\beta\sim U(0,\pi/2)$, and optimized using COBYLA for at most $80$ iterations. The optimization objective is the exact expectation value obtained from the statevector. After optimization, $1024$ samples are drawn from the final state distribution and the lowest-energy sampled bit string is returned as the candidate solution.

\subsection{Evaluation Metrics}
\label{subsec:metrics}

The evaluation considers both solution quality and computational characteristics. Solution quality is measured using feasibility and optimality gap. A solution is feasible when every customer is served exactly once and all selected missions satisfy the applicable payload and energy constraints. Coverage and physical feasibility are verified after decoding the binary solution.

The optimality gap is computed relative to the \gls{milp} reference objective as
\begin{equation}
    \mathrm{Gap} =
    \frac{C-C^{\star}}{C^{\star}}\times 100\%,
    \label{eq:optimality-gap}
\end{equation}
where $C$ is the objective value returned by the evaluated solver and $C^{\star}$ is the reference \gls{milp} objective. Optimality gaps are reported only for decoded solutions that satisfy both coverage and physical feasibility.

The structural complexity of the binary formulation is characterized by the number of \gls{qubo} variables and the number of nonzero quadratic interactions. These metrics quantify the effect of classical preprocessing on both the dimension and connectivity of the optimization problem.

Computational performance is evaluated using solver runtime. Solver runtime is recorded separately from candidate-route generation and \gls{qubo} construction time so that optimization performance is not conflated with preprocessing overhead. For stochastic methods, repeated runs are performed where specified by the experimental protocol, and aggregate statistics are reported. Runs skipped because the corresponding \gls{qubo} exceeds the configured \gls{qaoa} simulation limit are excluded rather than being treated as zero-runtime observations.

\subsection{Experimental Procedure}
\label{subsec:procedure}

For each problem size, $20$ independently generated synthetic delivery instances are evaluated using fixed random seeds. The same delivery instances are used across the compared solver configurations to ensure that differences between methods are not caused by different underlying problem realizations.

RQ1 uses customer counts of $4$, $5$, and $6$. For each instance, \gls{milp} and exact \gls{qubo} are evaluated once, while \gls{sa} and \gls{qaoa} are each repeated $10$ times using different solver seeds.

RQ2 uses customer counts of $4$, $5$, $6$, and $8$ and compares three preprocessing configurations: no filtering, payload-only filtering, and combined payload-and-energy filtering. For each instance and preprocessing configuration, one \gls{sa} run is performed. The reference objective is obtained from the combined payload-and-energy configuration.

RQ3 uses customer counts of $4$, $6$, $8$, $10$, and $12$ and evaluates one \gls{milp} run, one \gls{sa} run, and one \gls{qaoa} attempt per instance. Candidate routes are not capped for instances with at most eight customers. For larger instances, the candidate set is limited to $\max(80,12n)$ routes, corresponding to caps of $120$ routes for $n=10$ and $144$ routes for $n=12$. QAOA runs whose \gls{qubo} dimension exceeds the $16$-variable statevector limit are skipped.

For every run, the implementation stores feasibility, objective value, optimality gap, solver runtime, number of selected missions, coverage violations, physical violations, number of \gls{qubo} variables, and number of quadratic interactions. Candidate-generation and \gls{qubo}-construction times are recorded separately. The main experimental and solver settings are summarized in Table~\ref{tab:experimental-configuration}.
\section{Results}
\label{sec:results}

This section presents the results according to the three research questions. RQ1 evaluates solution quality, RQ2 examines the effect of classical preprocessing, and RQ3 analyzes computational scaling.

\subsection{RQ1: Solution Quality}
\label{subsec:rq1}

\begin{figure}[t]
    \centering
    \includegraphics[width=0.7\linewidth]{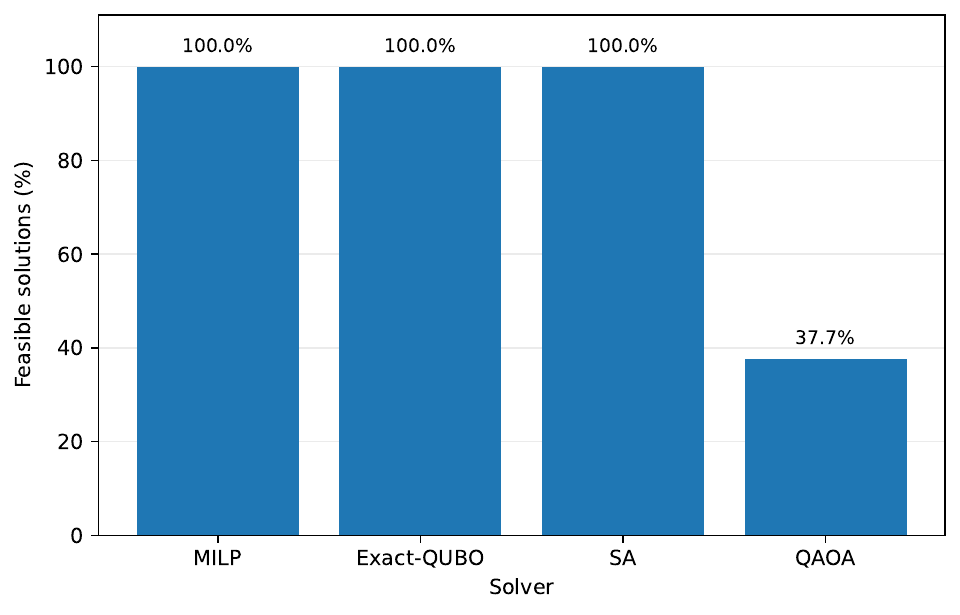}
    \caption{Feasibility rate of the compared solvers for RQ1.}
    \label{fig:rq1-feasibility}
\end{figure}

The exact \gls{qubo} formulation reproduced the \gls{milp} reference solution on all evaluated small instances, confirming that the binary reformulation preserves the optimum. Both the exact \gls{qubo} solver and \gls{sa} produced feasible solutions in $100\%$ of the runs, whereas \gls{qaoa} achieved a feasibility rate of $37.67\%$. Figure~\ref{fig:rq1-feasibility} summarizes these results.

\begin{figure}[t]
    \centering
    \includegraphics[width=0.7\linewidth]{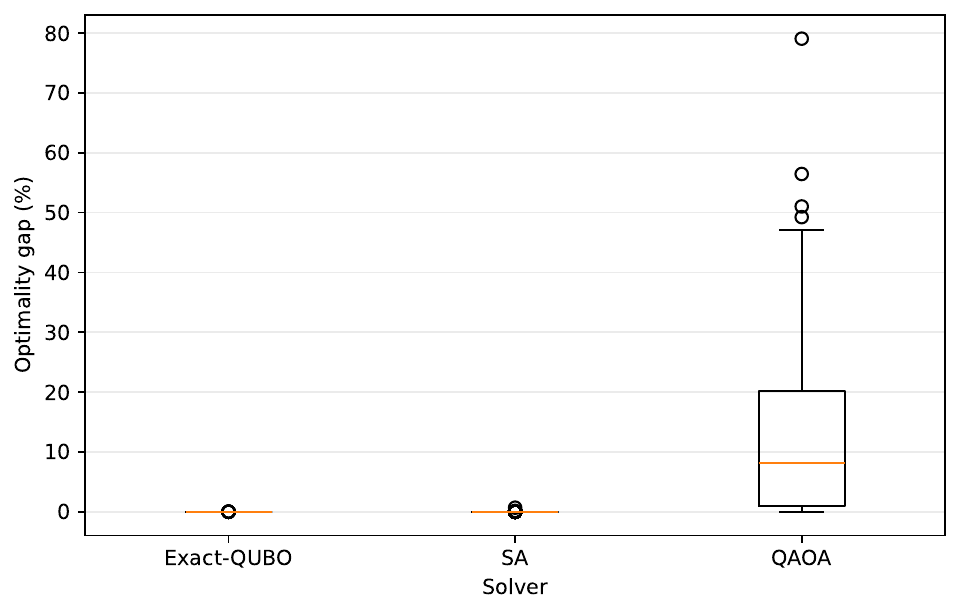}
    \caption{Optimality-gap distribution for the \gls{qubo}-based solvers. Gaps for \gls{qaoa} are reported for feasible decoded solutions.}
    \label{fig:rq1-gap}
\end{figure}

The solution-quality comparison shows a similarly clear difference. \Gls{sa} achieved a median optimality gap of $0\%$ and a mean gap of only $0.0014\%$. Among feasible \gls{qaoa} solutions, the median gap was $8.09\%$ and the mean gap was $12.36\%$. The corresponding distributions are shown in Fig.~\ref{fig:rq1-gap}. These results indicate that the considered \gls{qubo} formulation can be solved accurately by exact enumeration and \gls{sa}, while the evaluated shallow \gls{qaoa} configuration has greater difficulty producing feasible and near-optimal solutions.

\FloatBarrier
\subsection{RQ2: Effect of Classical Preprocessing}
\label{subsec:rq2}

\begin{figure}[t]
    \centering
    \includegraphics[width=0.7\linewidth]{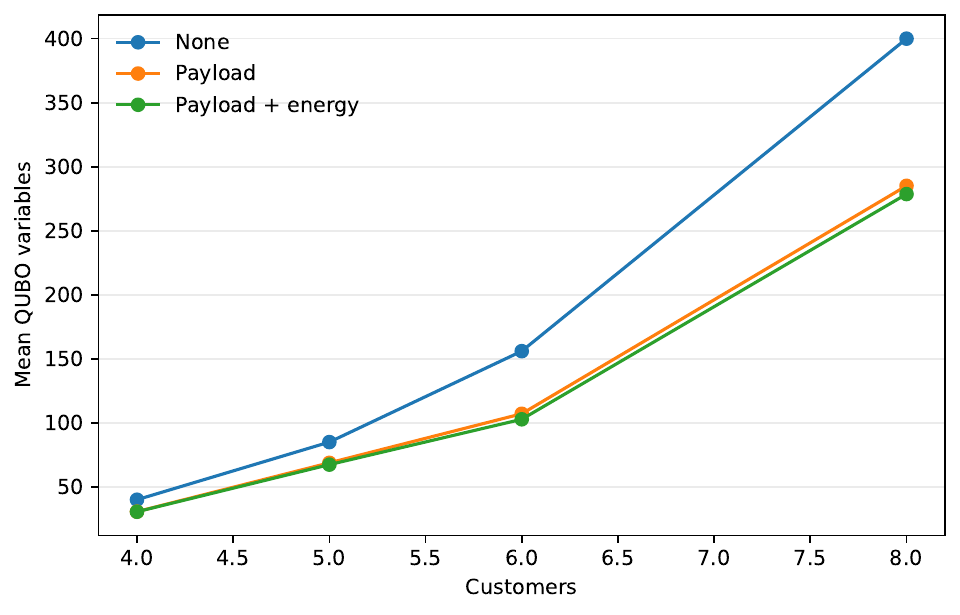}
    \caption{Mean number of \gls{qubo} variables under the three preprocessing configurations.}
    \label{fig:rq2-variables}
\end{figure}

Classical preprocessing substantially reduced the size of the resulting \gls{qubo}. Without filtering, the formulations contained $170.25$ variables on average. Combined payload-and-energy filtering reduced this value to $119.86$, corresponding to a reduction of $29.60\%$. Figure~\ref{fig:rq2-variables} shows that the difference becomes more pronounced as the number of customers increases.

\begin{figure}[t]
    \centering
    \includegraphics[width=0.7\linewidth]{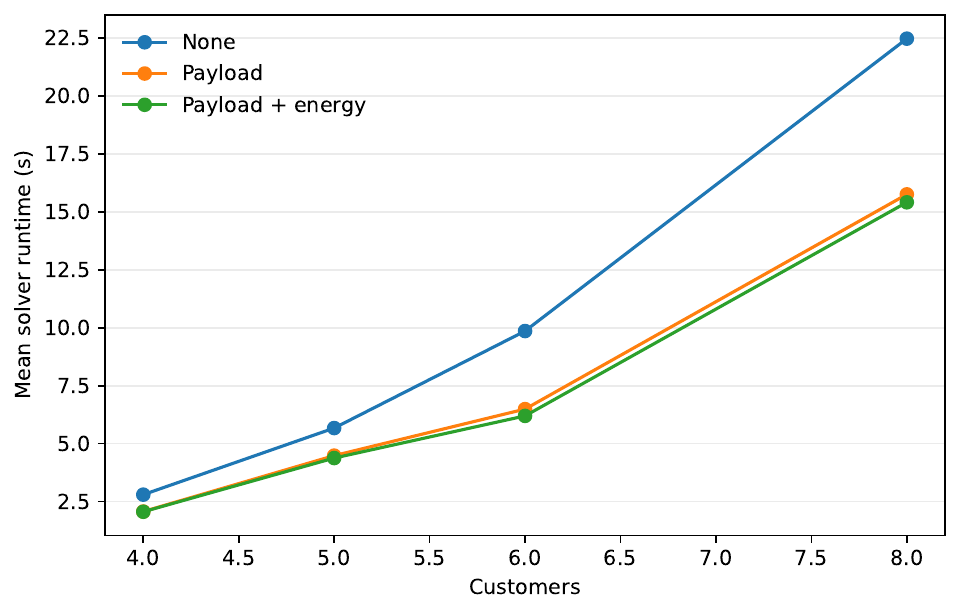}
    \caption{Mean \gls{sa} runtime for different preprocessing configurations.}
    \label{fig:rq2-runtime}
\end{figure}

The reduction in quadratic interactions was even larger, reaching $49.20\%$ for combined payload-and-energy filtering. This decrease in model size was accompanied by a $31.26\%$ reduction in mean \gls{sa} runtime. At the same time, feasibility increased from $66.25\%$ without filtering to $100\%$ with combined filtering. The runtime behavior is shown in Fig.~\ref{fig:rq2-runtime}. Importantly, the mission combination corresponding to the reference \gls{milp} optimum remained available after preprocessing on all evaluated instances.

These results support the use of classical feasibility checks before construction of the \gls{qubo}: the preprocessing stage reduces both the dimensionality and connectivity of the binary optimization problem without degrading the best-known solution on the tested instances.

\subsection{RQ3: Computational Scaling}
\label{subsec:rq3}

\begin{figure}[t]
    \centering
    \includegraphics[width=0.7\linewidth]{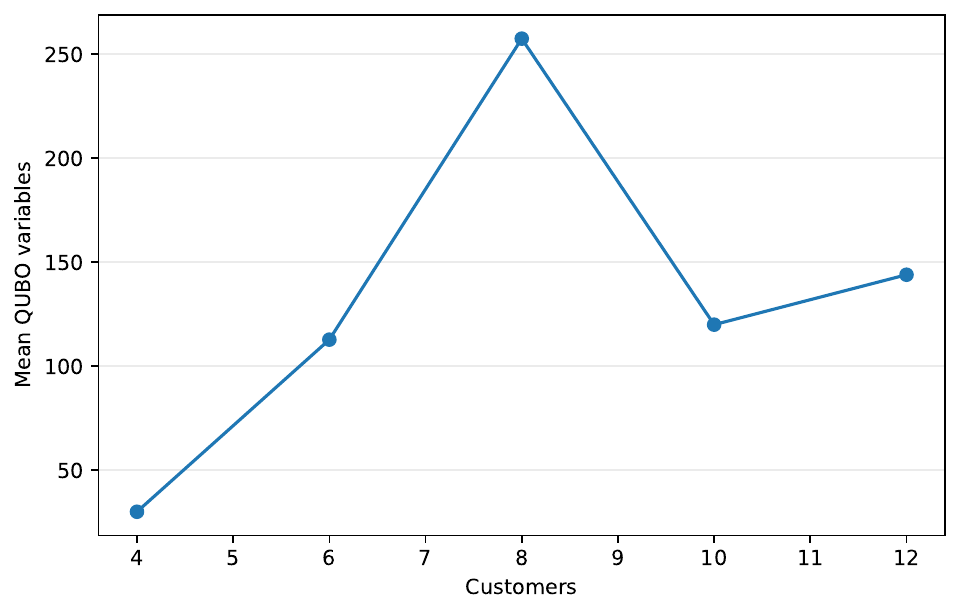}
    \caption{Mean \gls{qubo} size as a function of the number of customers. The decrease after eight customers results from candidate-route capping.}
    \label{fig:rq3-qubo-size}
\end{figure}

The scaling results show that solver runtime is more closely related to the resulting \gls{qubo} size than to customer count alone. As shown in Fig.~\ref{fig:rq3-qubo-size}, the mean \gls{qubo} size increases up to the eight-customer configuration, after which the candidate-route cap reduces the number of retained binary variables. Consequently, customer count does not grow monotonically with \gls{qubo} dimension.

\begin{figure}[t]
    \centering
    \includegraphics[width=0.7\linewidth]{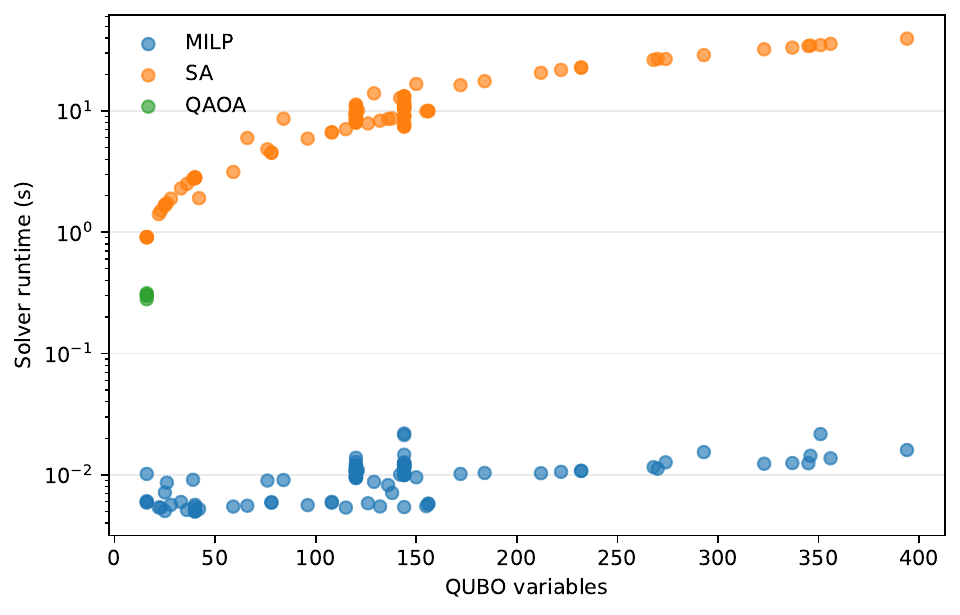}
    \caption{Solver runtime as a function of the number of \gls{qubo} variables. Only successfully executed runs are shown.}
    \label{fig:rq3-runtime-qubo}
\end{figure}

Figure~\ref{fig:rq3-runtime-qubo} therefore reports solver runtime directly against the number of \gls{qubo} variables. \Gls{sa} runtime increases clearly with the binary problem size, whereas \gls{milp} remains substantially faster over the tested range. The overall median runtime was $9.098$ s for \gls{sa} and $0.01003$ s for \gls{milp}. Only $4\%$ of the requested RQ3 \gls{qaoa} runs were executed because most formulations exceeded the configured statevector simulation limit. Skipped \gls{qaoa} runs are therefore excluded from the runtime analysis.

The results therefore indicate that the effective binary problem dimension, rather than the number of customers alone, is the more informative measure of computational difficulty in the proposed workflow. They also show that the current statevector-based \gls{qaoa} implementation is restricted to small \gls{qubo} instances and should not be interpreted as a large-scale solver.
\section{Discussion}
\label{sec:discussion}

The results show that the proposed formulation is technically valid but that the main benefit of the hybrid approach currently comes from classical problem reduction rather than from the quantum solver itself. The exact \gls{qubo} solution reproduces the \gls{milp} reference optimum on the evaluated small instances, which confirms that the mission-selection problem is preserved by the binary reformulation. At the same time, \gls{sa} consistently obtains feasible and near-optimal solutions, whereas the evaluated shallow \gls{qaoa} configuration produces feasible solutions substantially less often and with larger optimality gaps.

The strongest effect is observed for classical preprocessing. Removing payload- and energy-infeasible missions reduces the number of \gls{qubo} variables and quadratic interactions while also lowering \gls{sa} runtime. Importantly, the mission combination corresponding to the reference \gls{milp} optimum remains available after preprocessing on the tested instances. This indicates that deterministic feasibility checks can remove a significant part of the search space without eliminating the best-known feasible solution, supporting the decomposition of the workflow into a classical feasibility stage followed by a combinatorial \gls{qubo} optimization stage.

The scaling results further show that customer count alone is not sufficient to characterize computational difficulty. Because candidate-route generation and route capping directly determine the number of retained binary variables, instances with more customers can produce smaller \glspl{qubo} than instances with fewer customers. The resulting \gls{qubo} dimension is therefore a more informative indicator of solver cost than the nominal number of delivery requests.

The \gls{qaoa} results should be interpreted within the limitations of the evaluated configuration. The experiments use a shallow circuit and statevector simulation, which restricts execution to small \gls{qubo} instances. The observed results therefore do not demonstrate a quantum computational advantage. Instead, they highlight the difficulty of solving constrained mission-selection problems with shallow variational circuits and motivate further investigation of improved circuit depths, parameter optimization strategies, alternative mixers, and quantum hardware implementations.

Overall, the experiments suggest that near-term hybrid quantum--classical approaches to drone-delivery optimization are most useful when the classical component is used aggressively to reduce the binary search space before quantum optimization is attempted. In the current setting, classical exact and heuristic methods remain more effective in both runtime and solution quality, while the \gls{qubo} formulation provides a common representation through which classical and quantum solvers can be compared under the same optimization model.

\section{Threats to Validity}
\label{sec:threats}

Several factors limit the interpretation and generalizability of the results. First, the experiments use synthetic delivery instances with uniformly sampled customer locations and package weights. This provides controlled and reproducible test cases, but it does not capture all characteristics of real delivery networks, such as geographic obstacles, no-fly zones, heterogeneous demand, weather conditions, or time-dependent operations.

Second, the energy model is intentionally simplified. Energy consumption depends on traveled distance and carried payload, which captures the main effect relevant to the proposed formulation, but it does not model detailed aerodynamic behavior, wind, acceleration, battery degradation, or vehicle-specific power characteristics. The reported feasibility results should therefore be interpreted as feasibility with respect to the implemented model rather than as a complete physical validation of drone operation.

Third, candidate missions are limited to a maximum number of customer stops and can be capped for larger instances. This controls the size of the resulting \gls{qubo} and makes the scaling experiments computationally manageable, but it also changes the relationship between customer count and search-space size. Consequently, the reported scaling behavior applies to the implemented candidate-generation strategy and should not be interpreted as the scaling of an unrestricted drone-routing problem.
\FloatBarrier
Finally, the quantum evaluation is limited to a shallow \gls{qaoa} configuration executed with statevector simulation. The experiments therefore exclude hardware noise but are restricted to small \gls{qubo} instances because of the exponential memory requirements of statevector simulation. Different circuit depths, parameter-optimization methods, mixers, or quantum hardware may lead to different results. For this reason, the presented comparison characterizes the evaluated configuration rather than establishing a general performance limit for \gls{qaoa}.

\section{Conclusion}

\label{sec:conclusion}

This work investigated a hybrid quantum--classical approach to energy-constrained drone-delivery optimization in which candidate missions are generated and filtered classically before a \gls{qubo}-based mission-selection stage. The evaluation addressed solution quality, the effect of classical preprocessing, and computational scaling across \gls{milp}, exact \gls{qubo} optimization, \gls{sa}, and \gls{qaoa}.

The results show that the \gls{qubo} formulation preserves the reference optimum on the evaluated small instances, while \gls{sa} consistently produces feasible and near-optimal solutions. Classical payload and energy filtering substantially reduces both the number of \gls{qubo} variables and the number of quadratic interactions, while also lowering solver runtime and preserving the best-known solution on the tested instances. In contrast, the shallow statevector-based \gls{qaoa} configuration achieves lower feasibility and larger optimality gaps and remains restricted to small problem sizes.

Overall, the experiments indicate that the practical value of the proposed hybrid architecture lies primarily in reducing the binary optimization problem before the quantum stage. The study therefore supports classical feasibility preprocessing as an important component of near-term hybrid optimization workflows for drone delivery. Future work will consider richer operational constraints, more realistic energy models, alternative \gls{qaoa} configurations, and evaluation on quantum hardware or larger hybrid quantum backends.

\paragraph{Replication Package.}
The implementation, experiment configuration, raw results, and analysis scripts used in this study will be made available at Zenodo \cite{illesova2026reducing}.

\printglossary
\bibliographystyle{splncs04}
\bibliography{reference}

\end{document}